# Pressure-induced concurrent amorphization and superconductivity in topological material $NbNiTe_5$

Lingxiao Zhao[1#], Yi Zhao[1#], Bangshuai Zhu[1], Qi Wang[1,2], Cuiying Pei[1], Juefei Wu[1], Jin-Ke Bao[3], Wen-He Jiao[4]*, Yanpeng Qi[1,2,5]*

1. School of Physical Science and Technology, ShanghaiTech University, Shanghai 201210, China.
2. ShanghaiTech Laboratory for Topological Physics, ShanghaiTech University, Shanghai 201210, China
3. School of Physics and Hangzhou Key Laboratory of Quantum Matters, Hangzhou Normal University, Hangzhou 311121, China
4. Key Laboratory of Quantum Precision Measurement of Zhejiang Province, School of Physics, Zhejiang University of Technology, Hangzhou 310023, China
5. Shanghai Key Laboratory of High-resolution Electron Microscopy, ShanghaiTech University, Shanghai 201210, China

# These authors contributed to this work equally.

* Correspondence should be addressed to Y.Q. (qiyp@shanghaitech.edu.cn) and W.J. (whjiao@zjut.edu.cn)

**Abstract**: We have systematically studied the structural and electronic properties of a topological material $NbNiTe_5$ under high pressure. The evolution of the normal state resistance shows a non-monotonic trend from 0.7 GPa to 5.1 GPa, in accordance with the second-order transition along the inter-layer direction observed in X-ray diffraction and Raman spectra. At around 10 GPa, the sample starts amorphization, which is concurrent with the emergence of superconductivity. Upon further compression, the structural disorder enhances and the superconducting transition becomes clearer, suggesting that the superconductivity is modulated by the degree of disorder in $NbNiTe_5$ under high pressure. Within 45.7 GPa, the superconducting transition temperature ($T_c$) slowly rises from 0.6 K at 9.5 GPa to 1.4 K at 45.7 GPa. Our findings extend the family of transition metal chalcogenide superconductors and shed new light on understanding superconductivity in disordered systems.

## Introduction

For superconductors, it has long been intriguing to research the coexistence between structural disorder and superconductivity[1-3], such as amorphous metals (alloys)[4-6], and high-temperature quenched disordered phases[7,8], since the structural disorder generally confines the electrons and interrupts the dissipation-less transport in superconductors. However, a classic counterexample is the element bismuth, which is barely superconducting ($T_c$=0.5mK) in its crystalline form[9], however, exhibiting a moderate $T_c$ ($T_c$>6K) in amorphous or granular form[10,11]. Despite the cases being relatively rare, several disorder-induced or enhanced superconductors inspires that an appropriate degree of disorder could enhance the superconductivity in the amorphous system. Recently, Zhao et al reported a crystalline-amorphous-crystalline (CAC) phase transition in the *p* orbital-element compound $In_2Te_5$ under high pressure[12]. Though superconductivity persists within the whole pressure range, the transition temperature $T_c$ abruptly rises and reaches a maximum when the sample enters the amorphous phase. The continuous modulation of the degree of disorder in $In_2Te_5$ with high pressure reveals a clear dependence of $T_c$ on disorder, which is related to the special "block-hinge" structural characterizations of $In_2Te_5$. Other similar cases include evaporated amorphous phase of the phase change materials $GeSb_2Te_4$, $Ge_2Sb_2Te_5$, etc. (GSTs)[13-15] or $Sb_2Se_3$[8], which exhibit complex evolution of superconductivity under high pressure. Therefore, the successful modulation of the degree of disorder in these systems illustrates that high pressure is a powerful tool to explore more potential materials for the study of the correlation between structural disorder and superconductivity. Besides, the pressure-induced structural transition [16-18] [19,20] can be related to exotic electronic properties, including metal-insulator transitions[7,21,22], superconductor-insulator transitions[14,23] and topological phase transitions[15].

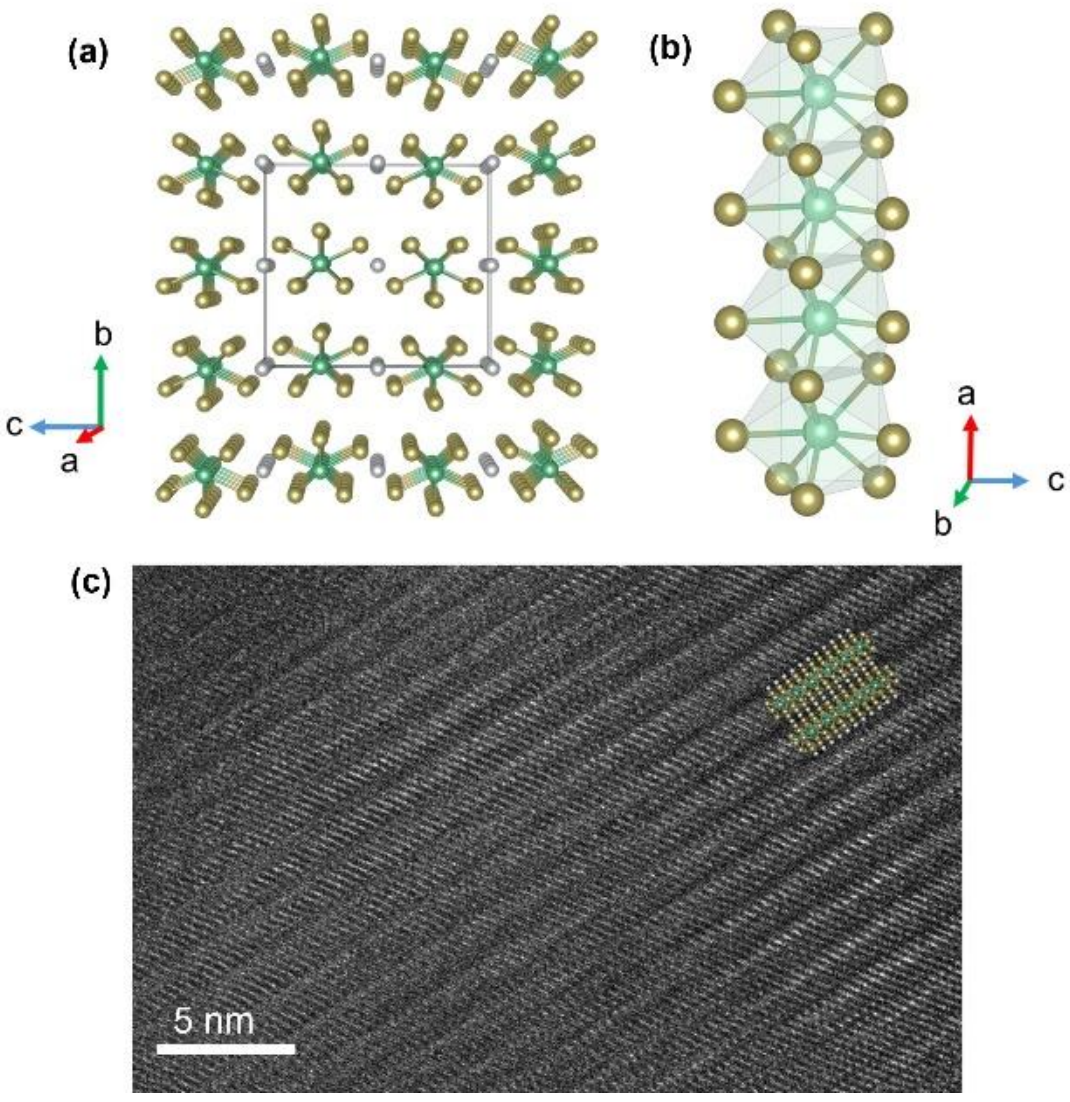

Figure 1. The crystal structure of $NbNiTe_5$. (a). The crystal structure from the (100) direction. The Nb, Ni, Te atoms are displayed with green, grey and brown solid balls. (b). The crystal structure from the (010) direction. (c). The high-resolution transmission electron microscope (HRTEM) image of $NbNiTe_5$ from the (010) direction. A schematic atomic structure is displayed in the image.

Pressure-induced superconductivity accompanied by amorphization is rare in general, especially in transition metal chalcogenides (TMC) or halogenides[24]. A typical example is the $(Ta/NbSe_4)_xI$ family[25,26]. In these materials, the Se atoms surround the Ta/Nb atom chains and become disordered under high pressure, while the metal chains retain long-range periodicity. This implies that a TMC with quasi-1D chain features is more likely to be a candidate with coexisting structural disorder and superconductivity. Hence, we concentrate on the recently reported layered TMC $NbNiTe_5$, a topological material with Dirac nodal lines[27]. $NbNiTe_5$ crystallizes in the space group *Cmcm* (Figure 1), consisting of a quasi-1D structure formed by Nb-Te chains and Ni atomic chains. Seen from the (100) direction, the local Nb-Te motif forms a distorted square antiprism by Te atoms. Moreover, the antiprisms extend in identical directions along the b axis, while in alternating directions along the c axis. The HRTEM image (Figure 1c) depicts the quasi-1D crystal structure along the (010) direction, which is similar to the *Cc* structure of $In_2Te_5$. Thus, $NbNiTe_5$ can possibly provide a platform for pressure-induced amorphization and relate it to superconductivity.

In this work, we have utilized high pressure to manipulate the structure and the electronic properties of $NbNiTe_5$. The transport results show a non-monotonic evolution of resistance ranging from 0.7 GPa to 5.1 GPa, which may be related to the second-order transition observed in Raman spectra and synchrotron X-ray diffraction (XRD). Upon further compression, superconductivity appears above 10 GPa, concurrently with the onset of amorphization. Unlike $Pd_3P_2S_8$[28] or several other examples[24,29,30], where superconductivity emerges at about 10 GPa prior to amorphization (>20 GPa), $NbNiTe_5$ exhibits concurrent onset of superconductivity and structural disorder (~10 GPa), with the XRD pattern exhibiting a broad diffuse scattering hump at around $2\theta = 13^\circ$ and Raman spectra retaining localized vibrational modes indicative of preserved short-range order. In addition, the relative Raman spectral weight at lower frequencies is significantly enhanced upon amorphization, suggesting a disorder-driven reconstruction or redistribution of the vibrational spectrum, which may be relevant to the emergence and enhancement of superconductivity. The relatively large pressure regime of gradual amorphization and enhancement of superconductivity makes $NbNiTe_5$ a rare case and a perfect platform for further detailed investigation of the mechanisms between disorder and superconductivity.

**Methods**

The high-quality $NbNiTe_5$ single crystals were synthesized using Te self-flux method, as described in Ref.[27]. The crystal structure is further confirmed by a HRTEM JEOL-F200. A piece of $NbNiTe_5$ single crystal is ground and transferred onto a copper grid with carbon membranes for the HRTEM observation. Temperature-dependent electrical transport property was measured in a cryostat, from 0.3 K to 2 K with the $^3$He module, and 1.8 K to 300 K with the conventional $^4$He module. A nonmagnetic diamond anvil cell (DAC) was employed to apply high pressure. A c-BN/epoxy mixture was utilized to insulate between BeCu gaskets and electrical leads and also serve as pressure transmitting medium (PTM) in transport experiments. Four Pt foils were arranged in a van der Pauw four-probe configuration to contact the sample in the chamber for resistance measurements. Pressure was determined by the ruby luminescence method[31]. The *in-situ* high-pressure Raman spectroscopy measurements at room temperature were performed using a Raman spectrometer (Renishaw inVia, U.K.) with a laser excitation wavelength of 532 nm. A symmetric DAC with anvil culet sizes of 300 μm (up to 50 GPa) was used with daphne oil 7373 as the pressure transmitting medium. The Raman spectra are all fitted by multiple Lorentzian functions. High-pressure synchrotron X-ray diffraction (XRD) measurements were carried out at room temperature with powder sample ground from single crystals at the beamline BL15U of Shanghai Synchrotron Radiation Facility (X-ray wavelength $\lambda = 0.6199$ Å). A symmetric DAC with anvil culet sizes of 300 μm (up to 50 GPa) and a T301 gasket were used. The two-dimensional diffraction images were analyzed using the Dioptas software. Le Bail refinements on crystal structures under high pressure were performed using GSAS-II[32].

The structure optimization and electronic structure calculations were carried out by the Vienna Ab initio Simulation Package (VASP) based on the density functional theory[33]. The cutoff energy of the plane-wave was set to 450 eV and the sampling grid spacing of the Brillouin zone was $2\pi \times 0.03$ Å$^{-1}$ in structure optimization[34]. The starting crystal structures were extracted from the high-pressure XRD results. The exchange-correlation functional is treated by the generalized gradient approximation of Perdew, Burke, and Ernzerhof[35]. The calculations use projector-augmented wave (PAW) approach to describe the core electrons and their effects on valence orbitals.

**Results and Discussions**

We utilize DAC to modulate the structural and electronic properties of $NbNiTe_5$ under high pressure. The temperature-dependent resistance is shown in Figure 2. In the temperature range of 1.8 K to 300 K, $NbNiTe_5$ shows a metallic transport behavior under high pressure, consistent with the reported results at ambient pressure. During compression, the resistance at room temperature first increases from 0.7 GPa to 1.7 GPa and then decreases from 1.7 GPa to 5.1 GPa. After 5.1 GPa, the resistance monotonically increases (Figure 2a and b). A similar non-monotonic evolution of resistance during compression near this pressure range (3-5 GPa) is reproduced in Run 3 (Figure S1). We can observe a similar tendency in the resistance at 10 K, and both the resistance at 300 K and 10 K under pressure are summarized in Figure 2c.

To examine the possible superconducting transition at lower temperature[36,37], we have utilized the $^3$He module of the cryostat to measure the resistance of $NbNiTe_5$ in the temperature range of 0.3 K to 1.5 K. The results are shown in Figure 2d-f. At about 9.5 GPa, we observe a drop in resistance at low temperature. The onset transition temperature becomes clearer, accompanied by a sharper transition width during compression. Meanwhile, the onset transition temperature increases from 0.6 K at 9.5 GPa to 1.4 K at 45.7 GPa. To further confirm the superconductivity of $NbNiTe_5$ under high pressure, we have measured the resistance of $NbNiTe_5$ at 31.0 GPa and 45.7 GPa under an external magnetic field. As shown in Figure 2e, the onset transition temperature decreases upon increasing the magnetic field, confirming that the drop of resistance at about 1.4 K is a superconducting transition. By choosing the temperature where the normal state resistance drops to 90% as the transition temperature $T_c$, we use a generalized Ginzburg-Landau (GL) formula

$$\mu_0 H_{c2}(T) = \mu_0 H_{c2}(0) \times \left(1 - (T/T_c)^2\right) / \left(1 + (T/T_c)^2\right)$$

to fit the upper critical field in Figure 2f. The fitting results show that the upper critical field $\mu_0 H_{c2}(0)$ is 0.76 T at 45.7 GPa and 0.62 T at 31.0 GPa, respectively.

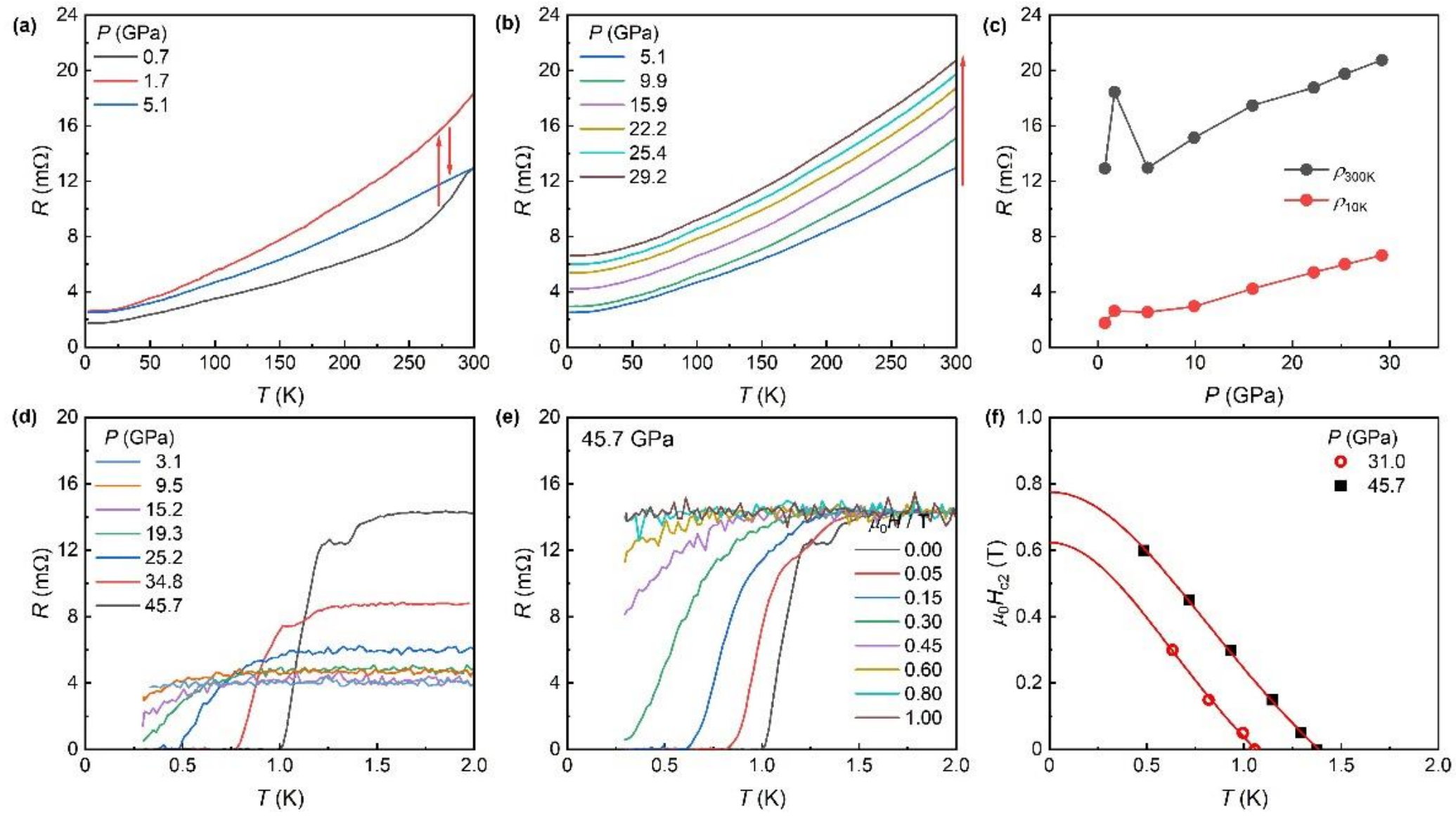

Figure 2. The evolution of resistance and superconductivity of $NbNiTe_5$ under high pressure. (a)-(b). The temperature dependence of resistance from 1.8 K to 300 K in the pressure range of 0.7 GPa and 29.2 GPa. The red arrows indicate the increase and decrease of resistance during compression. (c). The evolution of resistance at 300 K and 10 K under high pressure. The grey vertical line marks 4.5 GPa, the transition pressure. (d). The temperature dependence of resistance from 0.3 K to 1.5 K. (e). The temperature dependence of resistance at 45.7 GPa under external magnetic field from 0 T to 1 T. (f). The $\mu_0 H_{c2}$ at different temperature at 31.0 GPa and 45.7 GPa, respectively. The solid lines represent the fits based on the general GL formula.

To further understand the emergence of superconductivity under pressure, we have performed high-pressure *in-situ* XRD and Raman spectra, as summarized in Figure 3. At 0.8 GPa, the XRD peaks can be well indexed with the ambient structure. During compression, all of the diffraction peaks gradually move to higher angles. We have fitted the XRD data with the Le Bail method to acquire the cell parameters, as shown in Figure 3b. Around 4.5 GPa, the lattice parameter a has an anomalous increase, the lattice parameter b has a relatively sudden drop, while the lattice parameter c persists around 14.9 Å. This indicates that the structure is more compressed within the inter-chain direction than the intra-chain directions. Then, we fit the measured pressure and cell volume with the second-order Birch-Murnaghan (BM) model (Figure 3c).

$$P(V)=\frac{3B_0}{2}\times\left[\left(\frac{V_0}{V}\right)^{\frac{7}{3}}-\left(\frac{V_0}{V}\right)^{\frac{5}{3}}\right]$$

It should be noted that the bulk modulus $B_0$ abruptly increases from 58 GPa to 113.9 GPa at 4.5 GPa, indicating a hardening behavior under high pressure, which is almost in line with the non-monotonic evolution of the normal state resistance (Figure 2a and c). We should note that if we fit the whole pressure range from 0-14 GPa with a single second-order or third-order BM model, the fitted results are rather poor, or obtaining unreasonable parameters ($B_0$'=33), as shown in Figure S2. When the sample is further compressed to 10.4 GPa, all of the indexed XRD peaks start to disperse and leave a broad hump at about 13.5° after 20.2 GPa, which persists to the highest pressure 48.1 GPa. Our XRD results indicate the pressure-induced amorphization of $NbNiTe_5$ under higher pressure. Alongside the transport measurements, we observe that a higher degree of structural disorder is accompanied by a sharper superconducting transition. This suggests that superconductivity in $NbNiTe_5$ is correlated to the structural disorder.

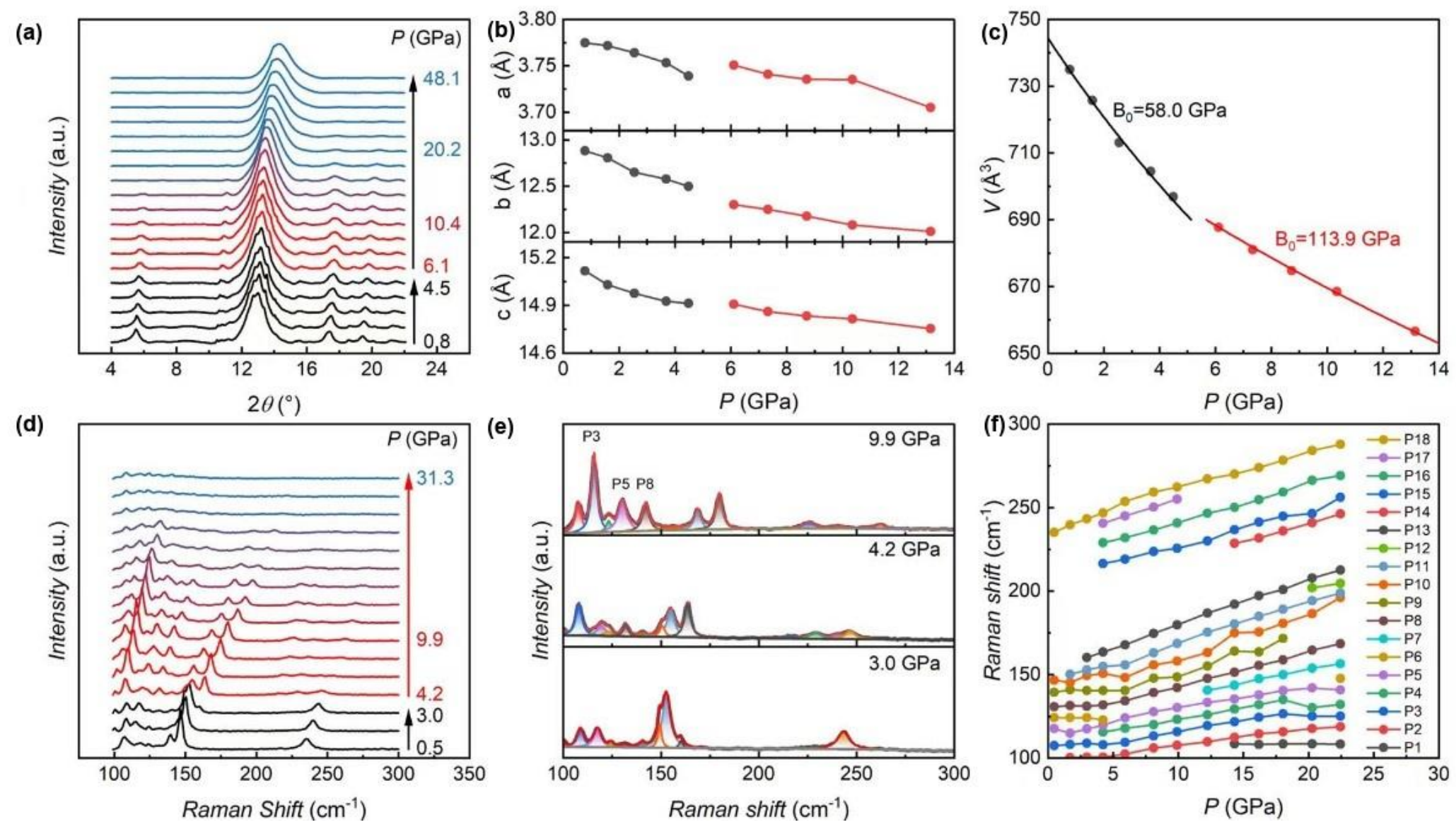

Figure 3. The high-pressure *in-situ* XRD and Raman spectra of $NbNiTe_5$. (a). The powder XRD pattern under pressure. (b). The evolution of cell parameters during compression. (c). The evolution of cell volume during compression. The black and red lines exhibit the fitted functions with the BM model. (d). The Raman spectra of $NbNiTe_5$ under pressure. (e). The fitted Raman spectra at 3.0 GPa, 4.2 GPa and 9.9 GPa with multiple Lorentzian functions. The vertical axes of the three panels represent identical scales of absolute intensity. (f). The evolution of Raman peaks during compression.

In addition to XRD, we have further carried out Raman spectra to unveil the evolution of the vibrational properties under high pressure. The pressure-dependent Raman spectra are shown in Figure 3d. When we compress the sample from 0.5 to 3.0 GPa, the main Raman peaks blueshift, which is a common behavior during compression.

Nevertheless, the overall Raman spectra undergo abrupt variations after 4.2 GPa (Figure S3). As plotted in Figure 3e and f, we summarize the fitted Raman spectra with multiple Lorentzian functions and choose the spectra at 3.0 GPa, 4.2 GPa and 9.9 GPa for illustrations. At about 4.5 GPa, we observe the emergence of new peaks at around 125 $cm^{-1}$ and 170 $cm^{-1}$, the enhancement of relative intensities of the peaks near 110 $cm^{-1}$, and the diminishing of the peaks around 250 $cm^{-1}$ (Figure 3e). Along with the XRD spectra and lattice parameters, we propose that the crystal structure of $NbNiTe_5$ keeps *Cmcm*, while the inter-chain strengthening modifies the bulk modulus and the intrinsic vibration modes. As for the pressure over 10 GPa (Figure 3f), we observe the emergence of new peaks labeled as P1, P7 and P14, and the relative intensities of the Raman peaks below 150 $cm^{-1}$ significantly enhance (P3, P5 and P8 in Figure 3e). Combined with the XRD results and the electronic transport properties measurements, we suggest that these changes in Raman spectra correspond to the gradual amorphization process. During the amorphization process, the long-range structural order decays while preserving the short-range motifs. Distortion of the local motif may appear, causing symmetry breaking and emergence of these new Raman modes we observed at about 10-15 GPa (Figure 3d and f). On the other hand, in disordered or amorphous systems, momentum selection rules are relaxed, and the measured Raman response can reflect a weighted contribution from vibrational states beyond the Brillouin-zone center, as an indirect indication of the vibrational density of states[38]. The enhancement of the relative intensities of the lower-frequency Raman modes indicates a redistribution of phonon (vibrational) density of states to lower energy. These significant changes in Raman spectra corresponding to the second-order transition (3-5 GPa) and the gradual amorphization (10-20 GPa) are reproduced in another run of measurements, as shown in Figure S5.

These two phenomena (anomaly in resistivity and superconductivity) during compression are related to changes in electronic and vibrational structures. First, we performed first-principles calculations on the electronic structures of $NbNiTe_5$ to clarify the non-monotonic evolution of the resistance at low pressure (Figure 2a). The calculated band structures and the density of states (DOS) at 1.6 GPa and 4.5 GPa are shown in Figure S6b and S6c. The lattice parameters are from the refinement of the high-pressure XRD results, and only the atom positions are free during the structural optimization. At both pressure points, multiple doubly degenerate bands form Dirac nodal lines in the Z-T-E-R-A plane in the reciprocal space (Figure S6a), similar to the electronic structure at ambient pressure[27]. Under high pressure, most bands shift upward under compression, leading to a slight change in the DOS (Figure S6b-c and Figure 4) near the Fermi energy, which may contribute to the non-monotonic evolution of resistance from 0.7 GPa to 5.1 GPa, jointly with some other possible mechanisms including significant changes in scattering processes.

Concurrent amorphization and superconductivity under compression are relatively rare among superconducting materials. Recent theoretical studies suggest that structural disorder may influence superconductivity not only through changes in the electronic background, but also through disorder- and anharmonicity-driven reconstruction of the vibrational spectrum. In particular, dissipative phonon scattering in disordered or amorphous solids can produce broadened, nonballistic vibrational excitations and redistribute vibrational spectral weight toward lower energies, thereby modifying the characteristic phonon energy scale and the effective electron-phonon coupling[39]. More recent Migdal-Eliashberg-based analyses further indicate that softened or strongly damped low-energy phonon modes can reshape $\alpha^2F(\omega)$ and, under suitable conditions, enhance $T_c$ [40-42].

In $NbNiTe_5$, the gradual amorphization between ~10 and 20 GPa coincides with the onset of superconductivity, while $T_c$ progressively increases upon further compression. Meanwhile, Raman spectra in this pressure range still retain localized vibrational features but show clear disorder-related mode reconstruction and a pronounced enhancement of the relatively low-frequency response below 150 cm$^{-1}$. Taken together, these observations suggest that pressure-induced amorphization in $NbNiTe_5$ is accompanied by a disorder-driven reconstruction of the low-energy vibrational spectrum, likely involving both spectral-weight transfer toward lower frequencies and enhanced damping/broadening of vibrational excitations. Such behavior is qualitatively consistent with recent theoretical proposals that low-energy softened or damped vibrational modes may strengthen electron-phonon-mediated pairing in disordered systems[39-42]. A quantitative determination of this connection, however, requires future high-pressure calculations and direct spectroscopic characterization of phonon linewidths and EPC.

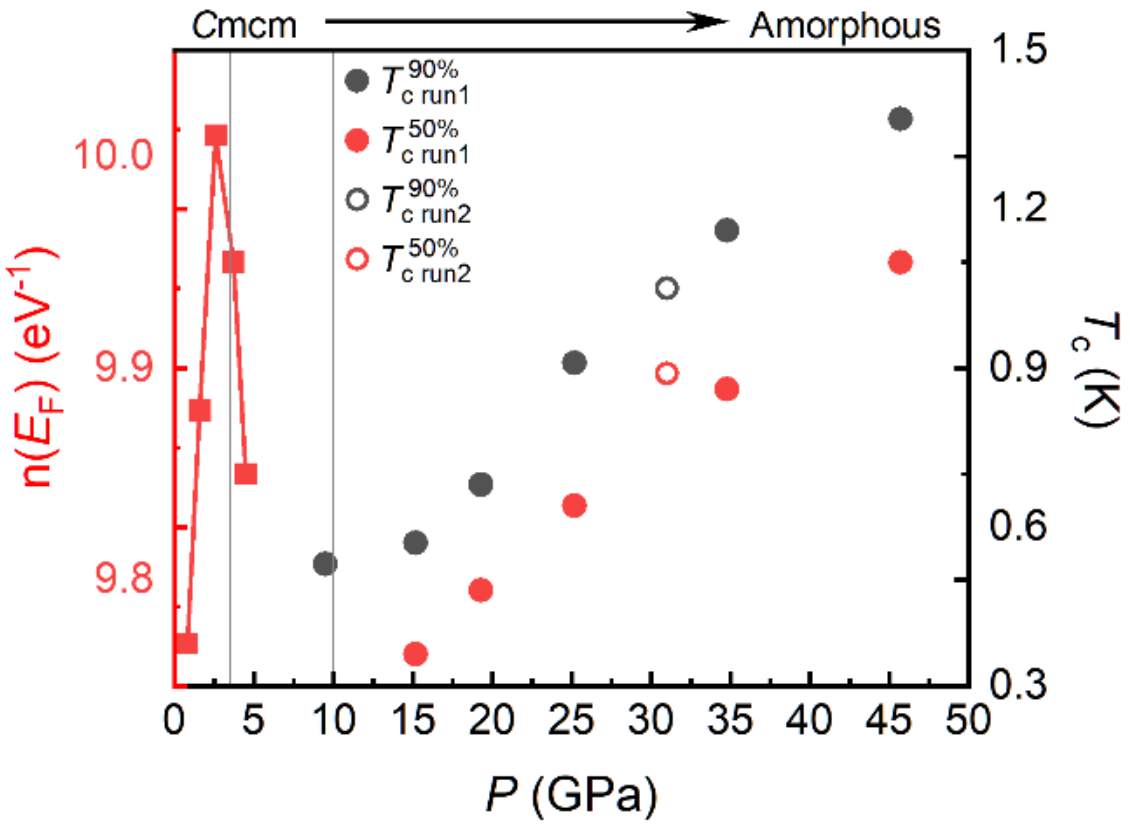

Figure 4. The phase diagram of superconductivity. The DOS is plotted on the left axis and the evolution of $T_c$ with 90% and 50% criteria is displayed with the right axis. The vertical grey lines note the hardening behavior and emergence of amorphization.

## Summary

In summary, we have systematically investigated the evolution of structural and electronic properties of $NbNiTe_5$ under high pressure through transport, XRD, Raman spectra and theoretical computations. The $NbNiTe_5$ crystal undergoes a non-monotonic evolution of resistance at 0.7 – 5.1 GPa, which is likely associated with the low-pressure second-order transition. At higher pressure, amorphization emerges and accompanies the onset of superconductivity. In the gradual amorphization process, $T_c$ is slowly enhanced, in accordance with the transitions observed in Raman spectra. As a rare case of concurrent pressure-induced amorphization and superconductivity in TMCs, this discovery suggests that similar compounds like $TaNiTe_5$, $TaPdTe_5$, etc. may exhibit similar structural evolutions under high pressure. The slow and gradual amorphization process upon compression provides an opportunity for precise and quantitative observation and analysis of the enhancement effect of disorder on superconductivity, which may inspire future research on the mechanism of the mysterious amorphous superconductivity.

**Acknowledgements**

This work was supported by the National Key R&D Program of China (Grant No. 2023YFA1607400), the National Natural Science Foundation of China (Grant No. 52272265). Wen-he Jiao acknowledges the Zhejiang Provincial Natural Science Foundation of China (Grant No. LZ23A040002) and the National Natural Science Foundation of China (Grant No. 11504329). Jin-ke Bao acknowledges National Natural Science Foundation of China (Grant No. 12204298). The authors thank the support from Analytical Instrumentation Center (# SPSTAIC10112914), SPST, ShanghaiTech University.

**Conflict of interest**

The authors declare that they have no conflict of interest.